\documentclass[3p,12pt]{elsarticle}

\usepackage{amssymb,amsmath,amsfonts,amstext,graphics,graphicx}

\newenvironment{definition}[1][Definition]{\begin{trivlist}
\item[\hskip \labelsep {\bfseries #1}]}{\end{trivlist}}

\def\p#1#2{\frac{\partial #1}{\partial #2}}

\newcommand{\pnq}[3]{\frac{\partial^{#3} #1}{\partial #2^{#3}}}

\newcommand{\de}[2]{\frac{d #1}{d #2}}

\def\de{\delta}

\def\R{{\rm I\kern-.1567em R}}
\def\bq{\begin{equation}}
\def\eq{\end{equation}}
\def\bqy{\begin{eqnarray}}
\def\eqy{\end{eqnarray}}

 \def\Xint#1{\mathchoice
{\XXint\displaystyle\textstyle{#1}}%
{\XXint\textstyle\scriptstyle{#1}}%
{\XXint\scriptstyle\scriptscriptstyle{#1}}%
{\XXint\scriptscriptstyle\scriptscriptstyle{#1}}%
\!\!\int}
\def\XXint#1#2#3{{\setbox0=\hbox{$#1{#2#3}{\int}$ }
\vcenter{\hbox{$#2#3$ }}\kern-.5\wd0}}

\def\dashint{\Xint-}

\def\intr{\int_{\R}\!}
\def\intp{\int_{\R_{+}}\!\!\!}

\def\dashintr{\Xint-_{\R}\!}
\def\dashintp{\Xint-_{\R_{+}}\!\!\!}

\def\calf{\mathcal{F}}

\def\calh{\mathcal{H}}

\def\call{\mathcal{L}}

\def\calp{\mathcal{P}}
\def\calq{\mathcal{Q}}

\def\R{\mathbb{R}}

\def\E{{\rm I\kern-.1567em E}}
\def\A{{\rm I\kern-.1567em A}}
\def\P{{\rm I\kern-.1567em P}}
\def\V{{\rm I\kern-.1567em V}}

\def\bq{\begin{equation}}
\def\eq{\end{equation}}

\def\Om{\Omega}

\journal{Physica D}

\begin{document}

\begin{frontmatter}

\title{Caldeira-Leggett Model, Landau Damping, and   the Vlasov-Poisson System\footnote{Dedicated to Steve Childress whose unique combination of mathematical and physical insight has been 
both an inspiration and a guiding light.}}
\author{George I. Hagstrom and P. J. Morrison}
\address{
Department of Physics and Institute for Fusion Studies, 
The University of Texas at Austin, Austin, TX 78712.}

\begin{abstract}
The Caldeira-Leggett Hamiltonian (Eq.~(\ref{hcl}) below) describes the interaction of a 
discrete harmonic oscillator with a continuous bath of harmonic oscillators. This system is 
a standard model of dissipation in macroscopic low temperature physics, and has applications to superconductors, quantum computing, and macroscopic quantum tunneling. The similarities between the Caldeira-Leggett model and the linearized Vlasov-Poisson equation are analyzed,  and it is shown that the damping in the Caldeira-Leggett model is analogous to that of Landau damping in plasmas  \cite{Landau}. An invertible  linear transformation  \cite{MP92,morrison00} is presented that converts solutions of the Caldeira-Leggett model into solutions of the linearized Vlasov-Poisson system.
 \end{abstract}


\end{frontmatter}

\section{Introduction}

In 1946 Landau \cite{Landau}  theoretically predicted the collisionless damping of the electric field in a plasma governed by the Vlasov-Poisson system.   This result has been of great importance in the field of  plasma physics, and indeed collisionless or continuum damping, as it is sometimes called,  occurs in a wide variety of kinetic and fluid plasma models that possesses  a continuous spectrum.  For example, such damping  occurs in the context of Alfven waves in magnetohydrodynamics   (see e.g.\ Chap.~10 of \cite{goed}) and has been proposed as a mechanism for plasma heating in response to electromagnetic waves.

Many other systems also undergo Landau damping, both inside and outside of plasma physics.    It is not surprising that Landau
damping exists in   stellar dynamics governed by the  Jeans equation \cite{binney} because this equation is of Vlasov type but with an attractive interaction potential.     In fact,  Landau damping occurs in collisionless kinetic theories with
a rather large class of potentials, and recently has been proven rigorously to exist in the nonlinear case  \cite{Villani,Villani2}.  Landau damping exists in the context of the fluid mechanics  of shear flow (see e.g.\  \cite{BM95,BM02} which contains a list of original sources over a period of more than  50 years) and the description of wind driven water waves.   It also appears in  multiphase media  \cite{Ryutov} and  has been established for systems containing large numbers of coupled oscillators, most notably the Kuramoto model. This has implications for biological models describing the synchronization or decoherence of the flashing of fireflies and chirping of crickets as well as other phenomenon in mathematical biology \cite{Biolandau}.

Another class of continuum systems involves the interaction of a discrete oscillator with  a continuous bath of oscillators.  In these systems the oscillator can be a particle or one mode of some field, and the bath often represents thermal fluctuations or radiation. One of the first detailed treatments  of such a system is due to Dirac \cite{Dirac1927}, but early on Van Kampen also used such a model to describe the emission and absorption of light by an atom \cite{VK51}.  The single wave model of plasma physics, which describes both beam plasma and laser plasma interaction physics \cite{Kaufman78,tennyson,evstati}, is also an example. The example of interest in this paper is the Caldeira-Leggett model \cite{CL-Original}.

The Caldeira-Leggett model  was invented in order to study quantum tunneling in the presence
of dissipation and the quantum limit of Brownian motion \cite{CL-long}. A model of this type was deemed necessary because quantum mechanics is incompatible with frictional forces.  However, the Caldeira-Leggett model is a Hamiltonian  system that exhibits dissipation by coupling to a continuum, i.e.,   it has Landau damping.   The Caldeira-Leggett Hamiltonian is the sum of the Hamiltonian of a classical harmonic oscillator, the Hamiltonian of continuous bath of harmonic oscillators, and a linear coupling term between the discrete and continuous degrees of freedom. The discrete degree of freedom corresponds to a macroscopic system and the bath of 
oscillators represent the environment. The coupling causes the discrete oscillator to damp by transference of energy to the continuum. 
This system has become a standard model for studying the physics of low temperature quantum 
systems, and it has numerous applications ranging from the
understanding of superconducting circuit elements to qubits in quantum computers \cite{CL-Phase}.

We analyze the classical Caldeira-Leggett model using a procedure analogous to that used by Landau to analyze the Vlasov-Poisson system  of plasma physics. Following Landau, the initial value problem can be solved using the Laplace transform and the rate of
decay can be derived in the weak damping limit. This paralleling of  Landau's original calculation
suggests a connection between this system and the Vlasov-Poission system.  In fact, we will show that both systems can be mapped into a  normal form that is common to  a  large class of infinite-dimensional Hamiltonian systems that have a continuous spectrum \cite{MP92,morrison00,BM02,morrison03}. 

The Caldeira-Leggett model, like all Hamiltonian systems in the class,  has a  continuous spectrum that is responsible for the damping through phase mixing (filamentation) and the Riemann-Lebesgue lemma. Because this structure is shared by a number of important physical systems, it is interesting
to determine the nature of their similarities.  It is well-known that the properties of linear ordinary differential equations are closely tied to the spectra of their time evolution
operators. In fact, for  given spectra there are a number of normal forms. Any linear finite-dimensional Hamiltonian system can be reduced to one of these  normal forms (ODEs) through an appropriate transformation, and in this sense the behavior of such systems is completely understood. The theory of normal forms for infinite-dimensional
Hamiltonian systems is not nearly as well-developed as that for  finite-dimensional systems, but for some systems much is known. For systems with continuous spectra, the analog of diagonalization is conversion into a multiplication operator. If the original system is $\dot{f}=\call f$,  then a transformation  $T$ such that $T\call T^{-1}$ is a multiplication operator would diagonalize the system. Any two systems that have the same normal form would thus  be equivalent  through some linear transformation.

This procedure has been performed for the linearized Vlasov-Poisson equation \cite{morrison00}, and when the spectrum is purely continuous the time evolution operator is equivalent to the multiplication operator $x$. This discovery led to the discovery of an entire class of transformations diagonalizing linear
infinite-dimensional Hamiltonian systems of a certain form \cite{morrison03}. In fact, it is always possible to perform such a transform in the special case of a bounded, self-adjoint operator \cite{reedsimon}. The operators dealt with here are usually unbounded and non-normal (even if they did exist in a Hilbert space), as is often the case when dealing with continuous Hamiltonian matter models. A precursor to the discovery of such transformations is existence of a complete basis of singular eigenfunctions of the original equation, a treatment that is common for systems with continuous spectra that dates back to Dirac \cite{Dirac1927}. In fact these methods have been developed in parallel within the field of plasma physics beginning with the work of Van Kampen \cite{VK55}  and within condensed matter physics through the work of Dirac and later Fano \cite{fano}. Caldeira and his collaborators developed a diagonalization method for the Caldeira-Leggett model \cite{CL-diag}, although they were primarily interested in the time evolution of the discrete degree of freedom and thus did not write down the full inverse of their transformation. In this paper we complete the treatment of the Caldeira-Leggett system. Then,  we note that the normal form is the same as that of the Vlasov-Poisson system and that the models are thus equivalent through the use of an integral transform.

Specifically, in Sec.~\ref{sec:CL} we review the Caldeira-Leggett model and then, in the spirit of Landau \cite{Landau},  present its Laplace transform solution in Sec.~\ref{sec:laplace}.  This is followed by obtaining the singular eigenfunctions, in the spirit of Van Kampen \cite{VK55} and Dirac \cite{Dirac1927}, and the invertible integral transform  akin to that of \cite{morrison00} for transforming to normal form.  In Sec.~\ref{sec:equiv} we show explicitly how the Caldeira-Leggett model is equivalent to a case of the linearized Vlasov-Poisson system.  Finally, in Sec.~\ref{sec:con} we conclude.

\section{Caldeira-Leggett Hamiltonian}
\label{sec:CL}

As noted above, the Caldeira-Leggett model is an infinite-dimensional Hamiltonian system describing the interaction of a discrete degree of freedom with an infinite continuum of modes \cite{CL-long}. The continuum is typically referred to as the environment. The Caldeira-Leggett model has the following Hamiltonian:
\begin{align}
H_{CL}[q,p;Q,P] &=\frac{\Omega}{2}P^2+\frac1{2} \left(\Omega+\intp dx\, \frac{f(x)^2}{2x}\right) Q^2
\nonumber\\
& + \intp dx  \, \left[\frac{x}{2}(p(x)^2+q(x)^2)+Q q(x) f(x)\right]\,,
\label{hcl}
\end{align}
which together with the Poisson bracket 
\bq
\{A,B\} =\left(\p{A}{Q}\p{B}{P}-\p{A}{P}\p{B}{Q}\right) 
 +\intp dx\left(\frac{\delta A}{\delta q}\frac{\delta B}{\delta p}
-\frac{\delta A}{\delta p}\frac{\delta B}{\delta q}\right) \label{hclbkt}
\eq
produces the equation of motion for observables in the form $\dot{F} =\{F,\calh\}$, where $F$ is any functional of the discrete, $(Q,P)$, and continuum, $(q,p)$, coordinates and momenta. Note,   it is assumed that $f(x)$ is chosen so that the integrals of (\ref{hcl}) exist. The coefficient of $Q^2$ includes a frequency shift term that is used to make the
Hamiltionian positive definite.  We take $p$ and $q$ to be functions on the positive real line, $\R_+$,  and $P$ and $Q$ to be real numbers. Hamilton's equations for the Caldeira-Leggett system are thus, 
\begin{align}
\dot{q}(x) &= xp(x) \\
\dot{p}(x) &= -xq(x)-Qf(x) \\
\dot{Q} &=  \Omega P \\
\dot{P} &= -\left(\Omega+\intp  dx\, \frac{f(x)^2}{2x}\right) Q - \intp dx\, q(x)f(x)\,.
\end{align}

This system was originally introduced by Caldeira and Leggett in 1981 \cite{CL-Original}. 
They initially considered a very massive harmonic oscillator coupled to a large number of light
harmonic oscillators with varying frequencies, and then studied the limit of the light oscillators becoming a continuous spectrum. The coupling causes $Q$ to decay to zero with time, and therefore the system
can be used to model dissipation. This makes it an ideal system to model the effects of
dissipation in quantum mechanics and especially quantum tunneling. It has been extensively studied
and is frequently mentioned in the condensed matter literature. There have been some controversies about the
physics of the damping and the physicality of the initial conditions \cite{CL-Phase}.  Connecting this system with plasma physics,  where much intuition has been developed over the years about wave-particle interaction, can 
help to improve the understanding of its behavior.  For example, a clear picture of filamentation can be viewed in the numerical work of \cite{heath}. 

Systems with continuous spectra   exhibit   phase mixing or filamentation.  A wide variety of systems have this property, and of course most famously the Vlasov-Poisson system for which Landau first discovered his damping. We will demonstrate explicitly that the damping mechanism of  the Caldeira-Leggett model is Landau damping. Furthermore we will derive a transformation that converts the Caldeira-Leggett model into the equation describing the time evolution of a single mode of the linearized Vlasov-Poisson system. Two copies of the Caldeira-Leggett model will be canonically equivalent to the time evolution of a pair of opposite $k$ Fourier modes of the linearized Vlasov-Poisson equation.

\section{The Landau Solution and Landau Damping}
\label{sec:laplace}

One of the classical calculations in plasma physics is the solution of the linearized
Vlasov-Poisson equation using the Laplace transform. This yields a formula for the solution of
the initial value problem and also facilitates the derivation of the damping rate for the electric
field. It is possible to do the same thing for the Caldeira-Leggett model. We begin with the set of
Hamilton's equations that were written down in the previous section and eliminate the two momenta to 
derive a pair of second order equations for the coordinates, 
\begin{align}
\ddot{q}(x) &= -x^2q(x)-Qxf(x) \\
\ddot{Q} &= -\Omega_c^2 Q-\Omega\intp dx\, f(x)q(x) \,,
\end{align}
where for convenience we  use  the corrected frequency,
\bq
 \Omega_c^2:= \Omega^2 +\Omega \intp dx\,  {f(x)^2}/{2x}\,.
 \label{omc}
 \eq
Defining  the Laplace transform  of the coordinates by
\begin{align}
\tilde{q}(x,s) &= \intp dt\,  q(x,t)e^{-st} \\
\tilde{Q}(s) &= \intp dt\,  Q(t)e^{-st}
\end{align}
results in the following set of algebraic equations:
\begin{align}
s^2\tilde{q}(x,s) &= -x^2\tilde{q}(x,s)-\tilde{Q}(s)xf(x)+sq(0,x)+\dot{q}(0,x) \\
s^2\tilde{Q}(s) &= -\Omega_c^2\tilde{Q}(s)-\Omega\intp dx\, \tilde{q}(x,s)f(x)+sQ(0)+\dot{Q}(0)\,,
\end{align}
which  can be easily solved for $\tilde{Q}(s)$,  
\begin{align}
\tilde{Q}(s) &= \left[ -\Omega\intp \, dx\, \frac{f(x)(sq(0,x)+\dot{q}(0,x))}{s^2+x^2}+sQ(0)+\dot{Q}(0)\right]
\nonumber\\
  &  \hspace{ 1.25 in} \div\,  \left[{s^2+\Omega_c^2-\Omega\intp dx\, \frac{xf(x)^2}{s^2+x^2}}\right]\,.
 \end{align}

The Laplace transform is inverted using the Mellin inversion formula,  
\begin{align}
Q(t) &= \frac{1}{2\pi i}\int_{\beta-i\infty}^{\beta+i\infty}\!\!\! ds\, \tilde{Q}(s)e^{st} \,,
\end{align}
where $\beta$ is any real number that ensures  $\tilde{Q}(s)$ is analytic for $Re(s)>\beta$. This integral is usually evaluted using
Cauchy's integral formula, whence asymptotically in the long-time limit the behavior  of the solution is given 
by the poles of $\tilde{Q}(s)$. Thus, the solution will be dominated by an exponentially decaying term arising from  the pole of $\tilde{Q}(s)$ closest to the real axis. We assume the closest pole is indeed close to the real axis and that there are no poles with a positive real part, i.e., that the solutions are stable. This is the weak damping limit.  As long as $f(x)$ is H\"{o}lder continuous, the poles of $\tilde{Q}(s)$ come from 
the zeros of the denominator, so we are interested in the roots of the equation
\bq
0 = s^2+\Omega_c^2-\Omega\intp dx\, \frac{xf(x)^2}{s^2+x^2}
 = s^2+\Omega_c^2-\Omega\intr dx\, \frac{f(|x|)_-^2}{2(x-is)} \,. 
\eq
Here $f(|x|)^2_-$ is the antisymmetric extension of $f(x^2)$ defined by  $f(|x|)^2_-=sgn(x)f(|x|)^2$.
Making  the substitution $\omega=is$,  yields the dispersion relation,  which in the limit   $\omega$ tends to  the real axis becomes 
\begin{align}
\omega^2-\Omega_c^2+\frac{\Omega}{2} 
\dashintr dx \, \frac{f(|\omega|)_-^2 }{x-\omega} 
+\frac{i\pi \Omega}{2} {f(|x|)_-^2} &= 0\,,
\end{align}
where $\dashint$ denotes the Cauchy principal value integral.  For quantities not yet integrated,  we will denote this by ${\bf PV}$.  This equation can be viewed as  the  dispersion relation in the weak damping limit. Let $\omega_c$ be a real solution to the real part of the above equation. Then let $\omega$ be a root of the
previous equation, assume $\gamma=Im(\omega)$ is small,  and
solve for $\gamma$ to first order; i.e.\   $0= 2i\omega_c\gamma + {i \pi \Omega}{f(|\omega_c|)_-^2}/2$  or 
\bq
\gamma = -\frac{\pi\Omega}{4|\omega_c|} {f(|\omega_c|)^2}\,. 
\eq

There are a large number of methods used to derive damping of $Q$ for this model. The standard approach is
to attempt to prove that after suitable approximations $Q$ satisfies the equation of motion of a damped
harmonic oscillator. The treatment here is almost identical to the method that was used to treat the Vlasov-Poisson equation by Landau, and agrees with
other derivations of the damping rate in the weak damping limit\cite{CL-Original}.


\section{Van Kampen modes:  diagonalization of  the Caldeira-Leggett model}
\label{sec:vk}

The Laplace transform method is just one way to treat the Vlasov equation. Another way is to write the solution as a superposition of a continuous spectrum of normal modes, a method  attributed to Van Kampen  \cite{VK55}.   Such modes of the Vlasov equation are called the Van Kampen modes, and we will see that  they exist for the Caldeira-Leggett model as well.  We formally calculate the Van Kampen modes for this system and use them to motivate the definition of an invertible integral transform,  akin to those of \cite{MP92,morrison00,BM02,morrison03}, that maps the Caldeira-Leggett model to action-angle variables, the  normal form for this Hamiltonian model.  The nature of the transformation depends on the coupling function $f(x)$. In the present treatment we will assume that $f(0)=0$,  but that $f$ does not vanish otherwise. We will also assume that the dispersion relation does not vanish anywhere. This excludes the possibility of discrete modes embedded in the continuous spectrum.  The case where the Caldeira-Leggett model possesses such  modes will be treated in
future work.  As stated above, the normal form of the Caldeira-Leggett  Hamiltonian will be seen to be equivalent to that for the Vlasov-Poisson system  through the  integral transformation introduced in \cite{MP92,morrison00}. 

The first step is to obtain a solution 
with time dependence $\exp({-iut})$ and derive equations for the amplitudes of a single mode $(q_u,p_u,Q_u,P_u)$.  To this end consider 
\begin{align}
i uq_u(x) &= -x p_u(x) \nonumber\\
i up_u(x) &= xq_u(x) +Q_uf(x) \nonumber\\
i uQ_u &= - \Omega P_u \nonumber\\
i uP_u &=  \left(\Omega +\intp dx\, \frac{f(x)^2}{2x}\right)Q_u+\intp dx\, q_u(x)f(x)\,.
\label{nmodeqs}
\end{align}
Note, although we use the subscript,  $u\in\R$ is a continuum label. Eliminating the momenta from Eqs.~(\ref{nmodeqs}) yields
\begin{align}
(u^2-x^2)q_u(x) &= Q_uxf(x)
\label{first} \\
(u^2-\Omega_c^2)Q_u &= \Omega\intp dx\, q_u(x)f(x)\,, 
\label{Q} 
\end{align}
where recall $\Omega_c$ is defined by (\ref{omc}).  Of these,  (\ref{first}) is solved following Van Kampen (a generalized function solution that dates to Dirac \cite{Dirac1927}) giving the  general form for $q_u$ 
\begin{align}
q_u(x) &= {\bf PV}\, \frac{Q_uxf(x)}{u^2-x^2}+C_uQ_u\delta(|u|-x)\,. 
\label{qu}
\end{align}
Substitution of (\ref{qu}) into (\ref{Q}) determines   $C_u$, 
\begin{align}
u^2-\Omega_c^2 &= \Omega \dashintp dx\, 
\frac{xf(x)^2}{u^2-x^2} 
+\Omega C_uf(|u|) \\
C_u &= \frac{{u^2-\Omega_c^2}}{\Omega f(|u|)}-\dashintr dx \, \frac{f(|x|)_-^2}{2(u-x)f(|u|)}\,.
\end{align}

Therefore we can specify an initial condition on the amplitudes $Q_u$ and compute the corresponding coordinates and momenta by an integral over the real line. 
Each mode oscillates with a different real frequency, with the expression for the
solution  given by 
\begin{align}
q(x,t) &= \dashintr du \,  \frac{Q_uxf(x)}{u^2-x^2} e^{-iut} +\intr du \, C_uQ_u\delta(|u|-x)  e^{-iut} \\
Q(t) &= \intr du \, Q_ue^{-iut}\,, 
\end{align}
were $Q_u$ acts an amplitude function that determines  which Van Kampen modes are excited. 

The Caldeira-Leggett model can be diagonalized and solved by making use of the integral transform alluded to above. Previously, Caldeira et al.\  \cite{CL-diag} derived a transformation  
that diagonalizes the Caldeira-Leggett model.  However, they were interested in solving for the evolution of the variable $Q$ and therefore did not attempt to write down the full
inverse of the operator (except in a special case where they made use of the evolution of the reservoir). We will extend their results by deriving the inverse map that  we will use   to establish the  equivalence with the Vlasov-Poisson system. 

In order to define the transform, we introduce 
a number of other important maps and introduce our notation. Extensive use will be made of the Hilbert transform, which is defined for a function $g(x)$ on $\R$  by 
\[
H[g](v) = \frac{1}{\pi}\dashintr dx\,\frac{g(x)}{x-v}\,.
\]
We also need  some  Hilbert transform identities \cite{king,morrison00}. Let  $g$, $g_1$, and $g_2$ be functions of  $x\in\R$ and suppose that all the expressions we write down are well defined, then the following hold:
\begin{align}
H[H[g]] &= -g \\
H[g_1H[g_2]+g_2H[g_1]] &= H[g_1]H[g_2]-g_1g_2 \\
H[vg] &= vH[g]+\frac{1}{\pi}\intr dx\,  g\,. 
\end{align}

\noindent Next we define two functions, $\epsilon_R$ and $\epsilon_I$ by 
\bq
\epsilon_I = \pi f(x)^2 \qquad {\rm and}\qquad
\epsilon_R =  2\frac{{x^2-\Omega_c^2}}{\Omega}+\pi H[f(|x|)_-^2]\,.
\eq
These together with    $|\epsilon|^2:=\epsilon_I^2+\epsilon_R^2$  are used  to define the following integral transforms:
\begin{definition}
For functions $h(x)$  on $\R_+$, the transform 
\[
T_+[h](u):=\epsilon_R h(|u|)+\epsilon_I H[h(|x|)](u)\,,
\]
while
\[
\widehat{T_+}[h](u):=\frac{\epsilon_R}{|\epsilon|^2}h(u)-\frac{\epsilon_I}{|\epsilon|^2}H[h(|x|)](u)\,,
\]
Related to the above transforms are two more transforms, 
\[
T_-[h](u):=\epsilon_R h(|u|)+\epsilon_I H[sgn(x)h(|x|)](u)\,,
\]
and
\[
\widehat{T_-}[h](u):=\frac{\epsilon_R}{|\epsilon|^2}h(u)-\frac{\epsilon_I}{|\epsilon|^2}H[sgn(x)h(|x|)](u)\,.
\]

\end{definition}

Using the  transform $T_+$ it is possible to write the map from the amplitudes  of the Van Kampen modes $Q_u$ to the functions $(q(x),Q)$.  To see this consider the expression for $q(x)$ in terms of the amplitude function
$Q_u$, and simplify it using the Hilbert transform as follows:  
\begin{align}
q(x) &=  \dashintr du\,  \frac{Q_uxf(x)}{u^2-x^2} +\intr  du \, C_uQ_u \delta(|u|-x)  \nonumber\\
&=  \dashintr du\, \frac{xf(x)Q_u}{2u}\left(\frac{1}{u-x}+\frac{1}{u+x}\right)+C_xQ_x+C_{-x}Q_{-x}\nonumber \\
&= \pi xf(x)\left(H\left[\frac{Q_u}{2u}\right](x)+H\left[\frac{Q_u}{2u}\right](-x)\right)+2C_x(Q_x+Q_{-x}) \,.
\label{qeq}
\end{align}

Next, decompose $Q_u$ into its symmetric and antisymmetric parts: $Q_u=Q_{+u}+Q_{-u}$ and observe that the antisymmetric parts vanish from both sides of  (\ref{qeq}), 
\begin{align}
q(x) &= \pi xf(x)H[ {Q_{+u}}/{u}](x)+2C_xQ_{+x} \nonumber \\
 &= \pi f(x)H[Q_{+u}]+\left(2\, \frac{{x^2-\Omega_c^2}}{\Omega f(x)}-\dashintr dx'\, \frac{f(|x'|)_-^2}{(x-x')f(x)}\right)Q_{+x} \,,
 \label{qx}
\end{align}
where the second line follows from  the third Hilbert transform identity combined with the fact that $Q_{u+}/u$ is antisymmetric and thus has a vanishing integral.   Now multiply both sides of (\ref{qx})  by $f(x)$ and find 
\begin{align}
f(x)q(x) &= \pi f(x)^2H[Q_{+u}]+\left(2\, \frac{{x^2-\Omega_c^2}}{\Omega}-\dashint_{\R}\!\!dx' \, \frac{f(|x'|)_-^2}{(x-x')}\right)Q_{+x} \nonumber\\
 &= \epsilon_I H[Q_{+u}]+\epsilon_{R}Q_{+x} \nonumber\\
 &= T_+[Q_{+u}]\,. 
 \end{align}
 Now we are set to define a transformation. 

\begin{definition}
Let $Q_{+u}$ be a function on  $\R_+$, then the map 
\[
I_c[Q_+]:=\left(\frac{1}{f(x)}T_+[Q_{u+}],2\intp du\, Q_{u+} \right)
\,.
\]
\end{definition}

\medskip

The map $I_c[Q_+]$,  a  map from the Van Kampen mode amplitudes  to the original dynamical variables,  has an inverse.
To see this note that $\epsilon_R=S+H[\epsilon_I]$,  where $S=2({x^2-\Omega_c^2})/{\Omega}$, and let $g$ be a function on $\R_+$. Then, 
 \[
 H[Sg(|x|)]=SH[g(|x|)]+\frac{4u}{\pi\Omega}\intp dx\,  g\,,
 \]
 where we have used our Hilbert transform identities to move the $x^2$ outside of the Hilbert transform of $g$. Using this,  consider   the following sequence of identities:
\begin{align}
\widehat{T_+}[T_+[g]] &= \frac{\epsilon_R}{|\epsilon|^2}(\epsilon_R g+\epsilon_IH[g])-\frac{\epsilon_I}{|\epsilon|^2}H[\epsilon_Rg+\epsilon_IH[g]] \nonumber \\
&= \frac{\epsilon_R^2}{|\epsilon|^2}g+\frac{\epsilon_R\epsilon_I}{|\epsilon|^2}H[g]-\frac{\epsilon_IS}{|\epsilon|^2}H[g]-\frac{\epsilon_I}{|\epsilon|^2}
H[H[\epsilon_I]g+\epsilon_IH[g]]-\frac{\epsilon_I}{|\epsilon|^2}\frac{4u}{\pi\Omega}\intp dx \, g \nonumber \\
&= \frac{\epsilon_R^2}{|\epsilon|^2}g+\frac{\epsilon_R\epsilon_I}{|\epsilon|^2}H[g]-\frac{\epsilon_IS}{|\epsilon|^2}H[g]-\frac{\epsilon_I}{|\epsilon|^2}(H[\epsilon_I]H[g]
-g\epsilon_I) -\frac{\epsilon_I}{|\epsilon|^2}\frac{4u}{\pi\Omega}\intp dx \,g \nonumber \\
&= g+\frac{\epsilon_R\epsilon_I}{|\epsilon|^2}H[g]-\frac{\epsilon_IS}{|\epsilon|^2}H[g]-\frac{\epsilon_I}{|\epsilon|^2}H[\epsilon_I]H[g]-\frac{\epsilon_I}{|\epsilon|^2}\frac{4u}{\pi\Omega}\intp dx\, g \nonumber \\
&= g+\frac{\epsilon_R\epsilon_I}{|\epsilon|^2}H[g]-\frac{\epsilon_I}{|\epsilon|^2}H[g](S+H[\epsilon_I]) -\frac{\epsilon_I}{|\epsilon|^2}\frac{4u}{\pi\Omega}\intp dx\, g \nonumber \\
&= g+\frac{\epsilon_R\epsilon_I}{|\epsilon|^2}H[g]-\frac{\epsilon_R\epsilon_I}{|\epsilon|^2}H[g]-\frac{\epsilon_I}{|\epsilon|^2}\frac{4u}{\pi\Omega}\intp dx \, g  \nonumber \\
&= g-\frac{\epsilon_I}{|\epsilon|^2}\frac{4u}{\pi\Omega} \intp dx\, g \,,
\end{align}
where in each step use has been made of the various identities above. 
Because the integral of $Q_+$ is equal to $Q/2$,  we can define the inverse of $I$ as follows:
\[
 \widehat{I_c}[q(x),Q]= \widehat{T_+}[f(x)q(x)]+\frac{2u}{\pi\Omega}\frac{\epsilon_I }{|\epsilon|^2} Q\,.
 \]

The above transform ignores the $(p,P)$ variables and only produces the symmetric part of the Van Kampen modes.  We  derive the other half  of the transformation from a mixed variable generating functional. To this end,  define $Q_+=\bar{Q}$ and $Q_-=\bar{P}$ and  rescale the coordinate part of the transformation by choosing $Q=2\int_{\R_+} \!du \, \bar{Q}\sqrt{{\epsilon_I}/{(\pi|\epsilon|^2)}}$ :
\begin{align}
\bar{Q} &= \sqrt{\frac{\pi|\epsilon|^2}{\epsilon_I}}\left(\hat{T_+}[f(x)q(x)]+\frac{2u}{\pi\Omega}\frac{\epsilon_I}{|\epsilon|^2}Q\right) = \widehat{I}[q(x),Q] \\
(Q,q(x)) &= \left(2\int_{\R_+} \!\! du\,  \bar{Q}\sqrt{\frac{\epsilon_I}{\pi|\epsilon|^2}}, \frac{1}{f(x)}T_+\left[\sqrt{\frac{\epsilon_I}{\pi|\epsilon|^2}}\bar{Q}\right]\right) 
= I[\bar{Q}]\,.
\end{align}

Then we introduce the mixed variable type-2 generating functional 
\[
\calf[q,Q,\bar{P}] = \int_{\R_+}\!\!du\,  \bar{P} \, \widehat{I}[q(x),Q] \,, 
\]
which  produces the transformations in the usual way:
\begin{align}
p(x) &= \frac{\delta\calf}{\delta q} = f(x)\, \widehat{T_+}^{\dag}\left[\sqrt{\frac{\pi|\epsilon|^2}{\epsilon_I}}\bar{P}\right] \\
P &= \frac{\delta\calf}{\delta Q} = \int_{\R_+}\!\!du\, \frac{2u\bar{P}}{\Omega}\sqrt{\frac{\epsilon_I}{\pi|\epsilon|^2}}\,.
\end{align}

Calculating the adjoint of $\widehat{T}$ simplifies the resulting expression for $p(x)$, viz., 
\begin{align}
p(x) &= \frac{1}{f(x)}T_-\left[\sqrt{\frac{\epsilon_I}{\pi|\epsilon|^2}}\bar{P}\right]\,.
\end{align}

Now, analogous to $I$ we define the operator $J[\bar{P}]=(p(x),P)$ by
\begin{align}
J[\bar{P}] &= \left(\frac{1}{f(x)}T_-\left[\sqrt{\frac{\epsilon_I}{\pi|\epsilon|^2}}\bar{P}\right],\int_{\R_+}du\frac{2u\bar{P}}{\Omega}\sqrt{\frac{\epsilon_I}{\pi|\epsilon|^2}}\right)\,,
\end{align}
which can be inverted through the use of the Hilbert transform identities.  Define $\bar{P}_c= \bar{P}\sqrt{{\pi|\epsilon|^2}/{\epsilon_I}}$, and consider the expression
\begin{align}
\frac{\epsilon_R}{|\epsilon|^2}p(x)& -\frac{\epsilon_I}{|\epsilon|^2}H\left[sgn(x)p(|x|)\right] 
\nonumber\\
&=\frac{\epsilon_R}{|\epsilon|^2}(\epsilon_IH[sgn(u)\bar{P}_c]+\epsilon_R\bar{P}_c)
 -\frac{\epsilon_I}{|\epsilon|^2}H[sgn(x)\epsilon_IH[sgn(u)\bar{P}_c]+sgn(x)\epsilon_R\bar{P}_c]\,. 
\end{align}

Paralleling  the method used to invert the map from $\bar{Q}$ to $(q,Q)$, we see a  difference
occurs when evaluating the term ${\epsilon_I}H[sgn(x)\epsilon_{R}\bar{P}]/{|\epsilon|^2}$,  i.e. 
\begin{align}
\frac{\epsilon_I}{|\epsilon|^2}H[sgn(u)u^2\bar{P}_c] &= x^2\frac{\epsilon_I}{|\epsilon|^2}H[sgn(u)\bar{P}_c] 
+\frac{2}{\pi} \frac{\epsilon_I}{|\epsilon|^2}\intp du\, u\bar{P}_c 
\nonumber\\
&= x^2\frac{\epsilon_I}{|\epsilon|^2}H[sgn(x)\bar{P}_c]+\frac{\Omega}{\pi} \frac{\epsilon_I}{|\epsilon|^2}P\,. 
\end{align}
With this expression we can directly use the inversion calculation for the $(q,Q)$ case to obtain 
the following expression for the full transformation:
\begin{align}
\bar{P} &=  \sqrt{\frac{\pi|\epsilon|^2}{\epsilon_I}}\left(\widehat{T_-}[f(x)p(x)]+\frac{2}{\pi} \frac{\epsilon_I }{|\epsilon|^2} P\right)\\
\bar{Q} &=  \sqrt{\frac{\pi|\epsilon|^2}{\epsilon_I}}\left(\widehat{T_+}[f(x)q(x)]+\frac{2u}{\pi \Omega}\frac{\epsilon_I}{|\epsilon|^2} Q\right)\,. 
\end{align}

Applying this transformation to Hamilton's equations yields the equations for a continuum of
harmonic oscillators. This can be seen directly for both $\bar{P}$ and $\bar{Q}$.
\begin{align}
\dot{\bar{Q}} &=  \sqrt{\frac{\pi|\epsilon|^2}{\epsilon_I}}\left(\widehat{T_+}[f(x)\dot{q}(x)]+\frac{2u}{\pi \Om} \frac{\epsilon_I}{|\epsilon|^2} \dot{Q}\right) \nonumber\\
&=  \sqrt{\frac{\pi|\epsilon|^2}{\epsilon_I}}\left(\widehat{T_+}[xf(x)p(x)]+\frac{2u}{\pi}   \frac{\epsilon_I}{|\epsilon|^2} P \right)\nonumber\\
&= \sqrt{\frac{\pi|\epsilon|^2}{\epsilon_I}}\left(\frac{\epsilon_R}{|\epsilon|^2} \, uf(u)p(u) -\frac{\epsilon_I}{|\epsilon|^2} \, H[|x|f(|x|)p(|x|)] +
\frac{2u}{\pi}   \frac{\epsilon_I}{|\epsilon|^2} P\right)\nonumber \\
&= \sqrt{\frac{\pi|\epsilon|^2}{\epsilon_I}}\left(\frac{\epsilon_R}{|\epsilon|^2}\,  uf(u)p(u) -\frac{\epsilon_I}{|\epsilon|^2}\, uH[sgn(x)f(|x|)p(|x|)] 
+\frac{2u}{\pi} \frac{\epsilon_I}{|\epsilon|^2} P\right)\nonumber\\
&= u\bar{P}\,.
\end{align}

Similarly, for $\bar{P}$, 
\begin{align}
\dot{\bar{P}} &=  \sqrt{\frac{\pi|\epsilon|^2}{\epsilon_I}}\left(\widehat{T_-}[f(x)\dot{p}(x)]+\frac{2}{\pi} \frac{\epsilon_I}{|\epsilon|^2} \dot{P}\right)\nonumber\\
&= \sqrt{\frac{\pi|\epsilon|^2}{\epsilon_I}}\left( \widehat{T_-}[-xf(x)q(x)-f(x)^2Q] - \frac{2\Omega_s}{\pi}\frac{\epsilon_I }{|\epsilon|^2} Q -\frac{2}{\pi} \frac{\epsilon_I}{|\epsilon|^2} \intp dx\, f(x)q(x)\right)\nonumber\\
&= \sqrt{\frac{\pi|\epsilon|^2}{\epsilon_I}}\left(-u \widehat{T_+}[f(x)q(x)] + \frac{2}{\pi} \frac{\epsilon_I}{|\epsilon|^2}\intp dx\, f(x)q(x) - \widehat{T_-}[f(x)^2]\, Q
\nonumber\right)\nonumber\\
&\hspace{2 in} 
-\frac{2\Omega_s }{\pi}\frac{\epsilon_I}{|\epsilon^2|} \, Q -\frac{2}{\pi}\frac{\epsilon_I}{|\epsilon|^2}\intp dx\, f(x)q(x)
\nonumber \\
&= \sqrt{\frac{\pi|\epsilon|^2}{\epsilon_I}}\left(-u \widehat{T_+}[f(x)q(x)] -\frac{\epsilon_R}{|\epsilon|^2}\, f(x)^2\, Q + \frac{\epsilon_I}{|\epsilon|^2}\, H[sgn(x)f(|x|)^2] \, Q
-\frac{2\Omega_s}{\pi}\frac{\epsilon_I}{|\epsilon|^2} \, Q\right)
\nonumber\\
&= \sqrt{\frac{\pi|\epsilon|^2}{\epsilon_I}}\left(-u \widehat{T_+}[f(x)q(x)]-\frac{2u^2}{\pi \Om} \frac{\epsilon_I}{|\epsilon|^2}\, Q\right)
\nonumber\\
&= -u\, \bar{Q}\,. 
\end{align}

Let this full map be called $I_f$ and consider $I_f$ as a map from  the  Banach space   $L_p\times L_p\times\R^2$, $p>1$, to the Banach space $L_p\times L_p$. The operator $I_f$ is a bounded linear
functional between these two spaces, because each term is either a multiplication operator that is
bounded on $L_p$, an $L_p$ function, or a bounded function multiplied by the Hilbert transform, 
which is another bounded operator. In order to establish the equivalence with the normal mode it is important to specify the phase space of the dynamical variables. Using this map we can simply choose each functional space to be $L_p$ and have a well defined map in each case. This map demonstrates how the Caldeira-Leggett model can be written as a superposition of a continuous spectrum of singular eigenmodes.

Because the transformation to the normal form was a canonical one,  the normal form  Hamiltonian  should be  the original Hamiltonian of the Caldeira-Leggett model written in the new coordinates. We will verify this by direct substitution. For convenience we introduce the quantities 
\begin{align}
A&=\frac{\Omega}{2}P^2+ \frac1{2} \int_{\R_+}\!\!dx\, {x} p(x)^2
\nonumber\\ 
B&=\frac{\Omega_s}{2}Q^2+\int_{\R_+}\left(\frac{x}{2}q(x)^2+f(x)q(x)Q\right)dx\,,
\nonumber
\end{align}
where $\Omega_s=\Omega_c^2/\Omega$.  Evidently,   $H_{CL}=A+B$. Then, 
\begin{align}
A &= \frac{\Omega}{2}P\int_{\R_+}\!\!du \,\frac{2u\bar{P}}{\Omega}\sqrt{\frac{\epsilon_I}{\pi|\epsilon|^2}}+\frac{1}{2}\int_{\R_+}\!\!du \, x p(x)f(x)\widehat{T_+}^{\dag}
\left[\sqrt{\frac{\pi|\epsilon|^2}{\epsilon_I}}\bar{P}\right] 
\nonumber\\
&=  P\int_{\R_+}\!\!du \, u\bar{P} \sqrt{\frac{\epsilon_I}{\pi|\epsilon|^2}}+\frac{1}{2}\int_{\R_+}\!\!du \, \widehat{T_+}\left[
x p(x)f(x)\right]\sqrt{\frac{\pi|\epsilon|^2}{\epsilon_I}}\bar{P}
\nonumber\\
&= P\int_{\R_+}\!\!du \, u\bar{P} 
\sqrt{\frac{\epsilon_I}{\pi|\epsilon|^2}}
+\frac{1}{2}\int_{\R_+}\!\!du \, u\widehat{T_-}\left[f(x)p(x)\right]
\sqrt{\frac{\pi|\epsilon|^2}{\epsilon_I}}\bar{P}
\nonumber \\
&= \frac1{2} \int_{\R_+}\!\!du \,{u}\bar{P}\left(\sqrt{\frac{\pi|\epsilon|^2}{\epsilon_I}}
\left(\widehat{T_-}[f(x)p(x)]+\frac{2}{\pi} \frac{\epsilon_I }{|\epsilon|^2} P\right)\right)
\nonumber\\
&= \frac1{2} \int_{\R_+}\!\!du \, u \bar{P}^2\,.
\label{ppart}
\end{align}

Similarly, 
\begin{align}
\frac{1}{2}\int_{\R_+}\!\!dx\, xq(x)^2 &= \frac{1}{2}\int_{\R_+}\!\!du\, 
\sqrt{\frac{\epsilon_I}{\pi|\epsilon|^2}}\widehat{T_+}^{\dag}\left[\frac{xq(x)}{f(x)}\right]\bar{Q}
\nonumber \\
 &= \frac{1}{2}\int_{\R_+}\!\!du\, \bar{Q}\sqrt{\frac{\epsilon_I}{\pi|\epsilon|^2}}
\left(u\frac{\epsilon_R q(u)}{f(u)} - uH\left[\frac{\epsilon_I(|x|)q(|x|)}{f(|x|)}\right]-2\int_{\R_+}\!\!dx\, f(x)q(x)\right)
\nonumber \\
&= -\frac{1}{2}\int_{\R_+}\!\!dx\, Q f(x)q(x) 
+\frac{1}{2}\int_{\R_+}\!\!du\,  \bar{Q}u\sqrt{\frac{\pi|\epsilon|^2}{\epsilon_I}}\widehat{T_+}\left[f(x)q(x)\right]\,.
\end{align}

Now, analyzing  the entire expression for $B$ in a sequence of steps, 
\begin{align}
B &= \frac{1}{2}\int_{\R_+}\!\!dx\, Qf(x)q(x)
+\frac{1}{2}\int_{\R_+}\!\!du\, \bar{Q}u\sqrt{\frac{\pi|\epsilon|^2}{\epsilon_I}}\widehat{T_+}\left[f(x)q(x)\right]
+ \frac{\Omega_s}{2}Q^2
\nonumber \\
&=\frac{1}{2}\int_{\R_+}\!\!du\, \bar{Q} u\sqrt{\frac{\pi|\epsilon|^2}{\epsilon_I}}\widehat{T_+}\left[f(x)q(x)\right]
+ \Omega_s Q  \int_{\R_+}\!\!du\,  \bar{Q}\sqrt{\frac{\epsilon_I}{\pi|\epsilon|^2}}
+\frac{Q}{2}\int_{\R_+}\!\!dx\, T_+\left[\sqrt{\frac{\epsilon_I}{\pi|\epsilon|^2}}\bar{Q}\right] 
\nonumber\\
&= \frac{1}{2}\int_{\R_+}\!\!du\, \bar{Q}u\sqrt{\frac{\pi|\epsilon|^2}{\epsilon_I}}\widehat{T_+}\left[f(x)q(x)\right]
\nonumber \\
& \hspace{.5 in} +\frac{Q}{2}\int_{\R_+}\!\!dx\,  \left(\left(\frac{2u^2}{\Omega}\bar{Q} 
+\pi H[sgn(x)f(|x|)^2]\bar{Q}\right)\sqrt{\frac{\epsilon_I}{\pi|\epsilon|^2}} 
+\epsilon_IH[\sqrt{\frac{\epsilon_I}{\pi|\epsilon|^2}}\bar{Q}]\right)
\nonumber \\
&= \frac{1}{2}\int_{\R_+} \!\!du\, \bar{Q}u\left(\sqrt{\frac{\pi|\epsilon|^2}{\epsilon_I}}\widehat{T_+}
\left[f(x)q(x)\right]+\frac{2u}{\Omega}\sqrt{\frac{\epsilon_I}{\pi|\epsilon|^2}} Q\right)
\nonumber \\
&=\frac1{2}\int_{\R_+}\!\!du\, u \bar{Q}^2\,.
\label{qpart}
\end{align}

With  (\ref{ppart}) and (\ref{qpart}) we obtain  $H_{CL}=\int_{\R_+}du {u} \left(\bar{Q}^2+\bar{P}^2\right)/2$ 
-- the Hamiltonian for a continuous spectrum of harmonic oscillators and
the normal form for the Caldeira-Leggett model.

\section{Equivalence to the Linearized Vlasov-Poisson Equation}
\label{sec:equiv}

The  treatment of the Caldeira-Leggett model of Sec.~\ref{sec:vk} is similar to an analysis of the linearized 
Vlasov-Poisson equation performed in  \cite{morrison00,MS08}. In those papers an integral 
transform was presented that transforms the Vlasov equation into a continuous spectrum 
of harmonic oscillators. The two systems are identical  except the spectrum of the Caldeira-Leggett model 
only  covers the positive real line.  Now we explicitly produce a transformation that takes one system into the other. 

The Vlasov equation describes the kinetic theory of a collisionless plasma. Spatially homogeneous
distribution functions are equilibria, and  linearization about such states are  
often studied in plasma physics. In the case of one spatial dimension and an equilibrium 
distribution function $f_0(v)$, the linearized Vlasov-Poisson equation around $f_0$ is given by 
\begin{align}
\p{f}{t}+v\p{f}{x}&-\frac{e}{m}\p{\phi}{x} f_0'  =0 \\
\pnq{\phi}{x}{2} &= -4\pi e \intr\!  dv\,  f 
\end{align}
where $f_0'=df_0/dv$.  These equations inherit the  noncanonical Hamiltonian structure of the full Vlasov-Poisson system \cite{morrison80} and have  a  Poisson bracket given by 
\bq
\{F,G\}_L =\int \!\int\! \! dxdv \, f_0\left[\frac{\de F}{\de f},\frac{\de G}{\de
f}\right]\,.  
\label{lvpb}
\eq
This bracket is of a  form that is typical for  Hamiltonian systems describing continuous media (cf.\ e.g.\ \cite{morrison82,morrison98}).  The Hamiltonian is given by
\bq
H_{L}=- \frac{m}{2} \int\int \! dv dx\,  v\, \frac{f^2}{f'_0}   
+ \frac{1}{8\pi}  \int \!dx\,  \left(\frac{\partial \phi}{\partial x}\right)^2 \,,
\label{KOH}
\eq
and the Vlasov-Poisson equation can be written as $\dot{f}=\{f,H_{L}\}_L$. 

The spatial dependence of the Vlasov-Poisson equation can be removed by performing a Fourier transform. 
This allows the potential to be explicitly eliminated from the equation
\begin{align}
\p{f_k}{t}-ikvf_k-\frac{4\pi i e^2}{mk}f_0'(v)\intr   dv\,  f_k &= 0\,.
\end{align}
The Hamiltonian structure in terms of the Fourier modes has the new bracket 
\bq
 \{F,G\}_L =\sum_{k=1}^{\infty}\frac{ik}{m} \intr   dv\,  f_0' \,
\left(\frac{\de F}{\de f_k}\frac{\de G}{\de f_{-k}}-
\frac{\de G}{\de f_{k}}\frac{\de F}{\de f_{-k}}\right) \,.
\label{klpb}
\eq
and the  Hamiltonian functional is simply  (\ref{KOH}) written in terms of the Fourier modes.  

One way to  canonize this bracket is with the following scalings:
\bq
q_{k}(v,t)=f_{k} \quad\quad{\rm and}\quad\quad  p_{k}(v,t)= \frac{mf_{-k}}{ikf_{0}'}\,,
\label{canvar}
\eq
where $k>0$. In terms of these variables the Poisson bracket has  canonical form,  i.e. 
\bq
 \{F,G\}_L =\sum_{k=1}^{\infty}  \intr  dv\,  
\left(\frac{\de F}{\de q_k}\frac{\de G}{\de p_{k}}-
\frac{\de G}{\de q_{k}}\frac{\de F}{\de p_{k}}\right) \,.
\label{klpb}
\eq

From this point it is possible to derive a canonical transformation that diagonalizes the Hamiltonian.
We make the following definitions:
\begin{align}
\varepsilon_I(v) &= -\frac{4\pi^2 e^2f_0'}{mk^2\intr dv\, f_0} \qquad\qquad\qquad\qquad
\varepsilon_R(v)   =  1+H[\varepsilon_I] \,,&\\
G_k[f] &= \varepsilon_Rf+\varepsilon_IH[f] \qquad\qquad \qquad\qquad
 \widehat{G_k}[f]   =  \frac{\varepsilon_R}{|\varepsilon|^2}f-\frac{\varepsilon_I}{|\varepsilon|^2}H[f]\,.&
\end{align}
It was proven in \cite{morrison00} that  $G_k=\widehat{G_k}^{-1}$.

A transformation to the new set of variables $(\calq_k,\calp_k)$ that diagonalizes the system will be given in terms of the   variables $(q_k,p_k)$.  To this end we first introduce the intermediate variables
$(\calq_k',\calp_k')$  defined by  
\bq
q_k = G_k[\calq_k'] \qquad{\rm and}  \qquad
\calq_k' = \widehat{G_k}[q_k]\,.
\eq
The corresponding momentum portion of the canonical transformation is induced by the following mixed variable generating functional:
\bq
\calf[q_k,\calp_k'] = \sum_{k=1}^{\infty}\intr du \, \calp_k'\widehat{G_k}[q_k]\,,
\eq
whence we obtain via $\calq_k'= {\de\calf}/{\de\calp_k'}$ and $p_k= {\de\calf}/{\de q_k}$, 
\bq
\calq_k' = \widehat{G_k}[q_k] \qquad{\rm and}  \qquad
\calp_k' = \widehat{G_k}^{\dag}[p_k]
\eq
Then,  the variables $(\calq_k,\calp_k)$ are defined as
\bq
\calq_k =  \left(\calq_k'-i\calp_k'\right)/\sqrt{2} \qquad{\rm and}  \qquad
\calp_k =  \left(\calp_k'-i\calq_k'\right) /\sqrt{2}\,, 
\eq
in terms of which the Vlasov-Poisson Hamiltonian has the form of a continuum of harmonic oscillators (see \cite{MS08} for an explicit calculation), 
\bq
H_L  = \sum_{k=1}^{\infty}\intr du \left(\calq_k^2+\calp_k^2\right)/2\,.
\eq

Thus, for a single value of $k$, this is the same normal form as that of the Caldeira-Leggett model, with the exception that the integral here is over the entire
real line instead of just the half line. If we consider two copies of the Caldeira-Leggett model the normal form would be the same as that for a single 
$k$ value of the linearized Vlasov-Poisson system. By composing the transformation that diagonalizes the Caldeira-Leggett model with the inverse of the transformation that diagonalizes  the  Vlasov-Poisson system we obtain 
a map that converts solutions of one system into solutions of the other system. 
Explicitly suppose that we have two copies of the Caldeira-Leggett Hamiltonian, with
the same coupling function $f(x)$. Then set the normal form of the second copy equal to the normal form of the Vlasov equation on the negative real line. 
Let $(q_1(x),p_1(x),Q_1,P_1)$ be one set of solutions to the Caldeira-Leggett model and let $(q_2(x),p_2(x),Q_2,P_2)$ be another and let $\Theta(x)$ be the Heaviside function. Then we can write a
solution to the linearized Vlasov-Poisson equation using the following map:
\begin{align}
f_k(v,t) &= G_k
\Bigg[
\frac{1}{\sqrt{2}}\Bigg(\Theta(u)\sqrt{\frac{\pi|\epsilon|^2}{\epsilon_I}} 
\left(\widehat{T_+}[f(x)q_1(x)]+\frac{2u}{\pi\Omega} \frac{\epsilon_I}{|\epsilon|^2}Q_1\right)
\\
& +\Theta(-u)\sqrt{\frac{\pi|\epsilon|(-u)^2}{\epsilon_I(-u)}}
\left(\widehat{T}_+[f(x)q_2(x)](-u)+\frac{-2u}{\pi\Omega}
\frac{\epsilon_I(-u)}{|\epsilon(-u)|^2}Q_2\right) \\
&+i\Theta(u)\sqrt{\frac{\pi|\epsilon^2|}{\epsilon_I}}
\left(\widehat{T_-}[f(x)p_1(x)]+\frac{2}{\pi}
\frac{\epsilon_I}{|\epsilon|^2}P_1\right) \\
&\  + i\Theta(-u)\sqrt{\frac{\pi|\epsilon(-u)|^2}{\epsilon_I(-u)}}
\left(\widehat{T_-}[f(x)p_2(x)]+\frac{2}{\pi}
\frac{\epsilon_I(-u)}{|\epsilon(-u)|^2}P_2\right)
\Bigg)
\Bigg] \\
f_{-k}(v,t) &= \frac{kf_0'}{m}G^{\dag}_k
\Bigg[
\frac{1}{\sqrt{2}}
\Bigg(\Theta(u)\sqrt{\frac{\pi|\epsilon|^2}{\epsilon_I}}
\left(\widehat{T}_+[f(x)q_1(x)]+\frac{2u}{\pi\Omega}
\frac{\epsilon_I}{|\epsilon|^2}Q_1\right)
\\
&+\Theta(-u)\sqrt{\frac{\pi|\epsilon|(-u)^2}{\epsilon_I(-u)}}
\left(\widehat{T}_+[f(x)q_2(x)](-u)+\frac{-2u}{\pi\Omega}
\frac{\epsilon_I(-u)}{|\epsilon(-u)|^2}Q_2\right)\\
&-i\Theta(u)\sqrt{\frac{\pi|\epsilon^2|}{\epsilon_I}}
\left(\widehat{T_-}[f(x)p_1(x)]+\frac{2}{\pi}
\frac{\epsilon_I}{|\epsilon|^2}P_1\right)\\
& -i\Theta(-u)\sqrt{\frac{\pi|\epsilon(-u)|^2}{\epsilon_I(-u)}}
\left(\widehat{T_-}[f(x)p_2(x)]+\frac{2}{\pi}
\frac{\epsilon_I(-u)}{|\epsilon(-u)|^2}P_2\right)
\Bigg)\Bigg] 
\end{align}

This map is invertible using the formulas presented earlier in the paper. Given a single mode of the linearized Vlasov-Poisson system,  $f_k(v,t)$,  we can
write two solutions to the Caldeira-Leggett model as follows:
\begin{align}
(q_1(x,t),Q_1(t)) &= I[\frac{1}{2}\Re(\hat{G}[f_k](u,t)+\hat{G}[f_k](-u,t))] \\
(p_1(x,t),P_1(t)) &= J[\frac{1}{2}\Im(\hat{G}[f_k](u,t)-\hat{G}[f_k](-u,t))] \\
(q_2(x,t),Q_2(t)) &= I[\frac{1}{2}\Im(\hat{G}[f_k](u,t)+\hat{G}[f_k](-u,t))] \\
(p_2(x,t),P_2(t)) &= J[\frac{1}{2}\Re(\hat{G}[f_k](-u,t)-\hat{G}[f_k](u,t))] 
\end{align}

Therefore one would expect the solutions of the Caldeira-Leggett model to share the same properties as the solutions of the Vlasov-Poisson system.

It was remarked earlier that both systems exhibit damping. In the Vlasov-Poisson case the electric field decays, and in the Caldeira-Leggett model it is the
discrete coordinate $Q$. The existence of the transformation between the two systems gives us a way to understand what determines the damping rate in each case.
In the standard calculation of the Landau damping rate for  the Vlasov equation,  it is clear that the rate depends only on the location of the closest zero in the lower half complex plane of the dispersion relation, which only depends on the equilibrium $f_0$. The same is true for the Caldeira-Leggett model, where the damping 
of $Q$ depends on the coupling function $f$. It is clear that integral transformations change the rate of damping, as all the instances of the Vlasov equation and
the Caldeira-Leggett model share the same normal form but generally have different damping rates. 

It is possible to interpret Landau damping using the normal forms and canonical transformation. The dynamical variables of the normal form have a time 
evolution $\sim \exp({-iut})$. The observables can then be expressed as some operator on this oscillatory dynamical variable. The result will be an oscillatory
integral over the real line, and by the Riemann-Lebesgue lemma we know that that such an integrated quantity will decay to zero in the long-time limit. For the systems at hand,   this integral can be deformed into the lower half complex plane, and Cauchy's theorem can be used to see that  the behavior is  governed  by
the locations of the poles of the analytic continuation of the oscillatory integrand. These poles determine the exponential damping rate. In these systems the poles are 
clearly introduced by the continuation (following the  Landau prescription) of   the dispersion relation in the integral transformations, which is therefore the origin of Landau damping. We will demonstrate  
this explicitly for  the damping of the coordinate $Q$ in the Caldeira-Leggett model.   

Starting  from the solution, 
\begin{align}
Q(t)&=\intr du \, (\bar{Q}(|u|)\cos(ut)+sgn(u)\bar{P}(|u|)\sin(ut))\frac{f(|u|)}{|\epsilon|} 
\nonumber\\
&=\intr du\, (\hat{I}[\mathring{q}(x),\mathring{Q}]\cos(ut)+\hat{J}[\mathring{p}(x),\mathring{P}]sgn(u)\cos(ut))\frac{f(|u|)}{|\epsilon|} \,, 
\label{Qsol}
\end{align}
we see that each term in the integrand of (\ref{Qsol}) has an oscillatory part and has poles at the zeros of $|\epsilon|^2$. The damping rate will be based on the closest zero of $|\epsilon|$,   the dispersion relation for the Caldeira-Leggett model. Likewise,  for  the Vlasov-Poisson system we can write a similar expression for the density $\rho_k(t)$, 
\begin{align}
\rho_k(t) &=\intr dv\, G_k[\hat{G}_k[\mathring{f}]e^{-iut}]
\nonumber\\
&=\intr du\, \left(\varepsilon_R(\hat{G}_k[\mathring{f}]e^{-iut})-H[\varepsilon_I]\hat{G}_k[\mathring{f}]e^{-iut}\right)
\nonumber\\
&=\intr du\, \hat{G}_k[\mathring{f}]e^{-iut}\,.
\end{align}
The  damping rates are given by the poles of $\hat{G}_k$, and the observed rate will be due to the closest zero of $|\varepsilon|^2$ to the real axis. Therefore, mathematically the source of the damping in the Vlasov-Poisson and Caldeira-Leggett models are identical,  it being  the nearest pole introduced by the integral transformation that diagonalizes the system.

\section{Conclusion}
\label{sec:con}
To summarize, we have shown how the  Caldeira-Leggett model can be analyzed  the same way as the Vlasov-Poisson system. We wrote down the solution using the Laplace transform,  an expression for the time evolution as an integral in the complex plane  over  the initial conditions.  It was then indicated how Cauchy's theorem can be used to derive the time asymptotic behavior of the solution,  and  it was described how the long-time damping rate is equal to the distance from the real axis of the closest  zero of the dispersion  relation  (when analytically continued into the lower half complex plane). Thus, the damping of  the Caldeira-Leggett model can be seen to be a rediscovery of  Landau (or continuum) damping.  Caldeira and Leggett introduced their  system to study damping in quantum mechanical systems, and it is now seen to be one of many interesting physical examples of Hamiltonian systems that exhibit such behavior.

Next we described  how to analyze the Caldeira-Leggett model by means of singular eigenmodes, paralleling  Van Kampen's  well-known treatment of the Vlasov-Poisson system.   Here the solution was written as an integral over a distribution of such modes, each of which is itself a solution that oscillates with some real frequency.  We described how  Hamiltonian systems with continuous spectra  generally have a solution formula in terms of such an integral over singular eigenmodes. This type of formal expansion  led to a an explicit integral transformation that transforms the original Caldeira-Leggett  system into a pure advection problem, just as is the case for  the Vlasov-Poisson system.   It was noted that a general class of such transformations  was written down in \cite{morrison00}  and was subsequently extended to a larger class of Hamiltonian systems \cite{morrison03}. The existence of these transformations amounts to a  theory of normal forms  for systems with a continuous spectrum, analogous to the theory of normal forms for finite degree-of-freedom Hamiltonian systems.  This enabled us to write  down an explicit transformation that converts the time evolution operator for the Caldeira-Leggett  Hamiltonian into a multiplication operator and we found the inverse of this map. In this way we showed that the Caldeira-Leggett model shares the same normal form as the Vlasov-Poisson system, along with  a number of other Hamiltonian systems that occur in different physical contexts. 

One reason for investigating Hamiltonian structure is the existence of universal behavior shared by such systems.  
For example,  linear Hamiltonian system with the same normal form are equivalent.  This suggests some further avenues for research.   Here we only treated the case where the dispersion relation of  the Caldeira-Leggett model has no roots with real frequency; i.e.\ spectrum was purely  continuous.  When there are roots, the spectrum is 
no longer purely continuous and there are embedded eigenvalues, as is known to be the case for the Vlasov-Poisson system \cite{Case}.   Consequently, one obtains a different normal form, one with a discrete component, and this  and more complicated normal forms  could be  explicated.  We expect that there is a transformation that takes  Vlasov-Poisson system with embedded modes into the Caldeira-Leggett model with embedded modes. 
Also, finite degree-of-freedom Hamiltonian systems are known to have only certain bifurcations of spectra, for example,  as governed by  Krein's theorem.  Since there is a generalization of this theorem  for Vlasov-like systems \cite{VlasovKrein}, one could investigate   bifurcations  in the context of the  Caldeira-Leggett model.  
Another possibility would be to use the tools developed  \cite{MS08}   to do statistical mechanics over the continuum bath.  Lastly, the integral transform we presented is intimately related to the Hilbert transform, which is known to be an important tool in signal processing.  In the same vein the integral transform for the Vlasov-Poisson system of Ref.~\cite{morrison00} has been shown to be a useful experimental tool \cite{skiff1,skiff2} and one could explore experimental ramifications in the context of the Caldeira-Leggett model. 

\section*{Acknowledgment}
\noindent This research was  supported by U.S. Dept.\ of Energy Contract \# DE-FG05-80ET-53088.

\bibliographystyle{elsarticle-num}

\bibliography{NegCaldeiraLeggett2}

\end{document}